\input psfig
\def\ltsim{\raise 2pt \hbox {$<$} \kern-1.1em \lower 4pt \hbox {$\sim$}}
\def\ltapprox{\raise 2pt \hbox {$<$} \kern-1.1em \lower 5pt \hbox {$\approx$}}
\def\gtsim{\raise 2pt \hbox {$>$} \kern-1.1em \lower 4pt \hbox {$\sim$}}
\def\gtapprox{\raise 2pt \hbox {$>$} \kern-1.1em \lower 5pt \hbox {$\approx$}}
\def\degrees{$^{\circ}$}

\documentclass[useAMS,usenatbib]{mn2e}

\title[Diffuse radio sources in Abell 548b]{Diffuse radio sources in 
the cluster of galaxies Abell 548b }
\author[l. Feretti et al.]
  {L.~Feretti,$^1$\thanks{email: lferetti@ira.inaf.it}
  M.~Bacchi,$^1$ O.B.~Slee,$^2$ G.~Giovannini,$^{3,1}$
  \newauthor 
  F.~Govoni,$^{1,4}$ H.~Andernach,$^5$
  G.~Tsarevsky,$^{2,6}$\\
$^1$ INAF Istituto di Radioastronomia, via P. Gobetti 101, I--40129
Bologna, Italy\\
$^2$Australia Telescope National Facility, CSIRO, P.O. Box 76, Epping,
NSW 1710, Australia\\
$^3$Dipartimento di Astronomia, Univ. Bologna, via Ranzani 1,
I-40127 Bologna, Italy\\
$^4$INAF Osservatorio Astronomico di Cagliari, Loc. Poggio dei Pini,
Strada 54, 09012 Capoterra, Italy\\
$^5$Departamento de Astronom\'{i}a, Universidad de Guanajuato, AP 144,
Guanajuato CP 36000, Mexico\\
$^6$Astro Space Center - RAS, 84/32 Profsoyuznaya St.,
Moscow, 117910 Russia
}
\date{MNRAS Submitted, }

\pagerange{\pageref{firstpage}--\pageref{lastpage}} \pubyear{}

\def\LaTeX{L\kern-.36em\raise.3ex\hbox{a}\kern-.15em
    T\kern-.1667em\lower.7ex\hbox{E}\kern-.125emX}

\begin{document}

\label{firstpage}

\maketitle

\begin{abstract}
We report extensive  VLA and ATCA observations
 of the two diffuse radio sources in the cluster of galaxies
Abell 548b, which confirm their classification as relics.  The two
relics (named A and B) show similar flux density, extent, shape,
polarization and spectral index and are located at projected distances of
about 430 and 500 kpc from the cluster center, on the same side of the
cluster's X-ray peak. On the basis of  spectral indices of
discrete radio sources embedded within the diffuse features, we have
attempted to distinguish emission peaks of the diffuse sources from
unrelated sources.  We have found that both relics, in particular the
B-relic, show possible fine structure, when observed at high
resolution. Another diffuse source (named C) is detected close in
projection to the cluster center.  High-resolution images show that it
contains two discrete radio sources and a diffuse component, which
might be a candidate for a small relic source.  The nature and
properties of the diffuse radio sources are discussed.  We conclude
that they are likely related to the merger activity in the cluster.
\end{abstract}

\begin{keywords}
Radio continuum: general -- 
Galaxies: clusters: general --
Galaxies: clusters: individual: A548b -- diffuse radiation. 
\end{keywords}

\section{Introduction}

In recent years growing interest has been directed toward the study of
diffuse radio sources in clusters of galaxies. 
Cluster radio halos have been found to  
permeate cluster centers, and diffuse radio sources classified
as relics have been detected in the cluster peripheries (see e.g. review by
Feretti 2003).  Extended relics, showing a steep
spectrum, high polarization degree and no obvious association with any
cluster galaxy, have been suggested to originate along cluster merger
shocks, either by diffusive shock acceleration of electrons
(En{\ss}lin et al. 1998) or adiabatic recompression of fossil radio
plasma (En{\ss}lin \& Br\"uggen 2002). Currently we know about 30
clusters showing relics, but detailed studies are available for only
a few sources of this class (see e.g. Giovannini \& Feretti 2004
for a review). In six clusters, the relics are double 
and located on symmetric sides with respect to the cluster
center. Relics may show quite different observational structures
and locations, which might indicate different physical properties.
New information on relic sources is therefore of great importance.

\begin{table*}
\caption{Observing log}
\begin{flushleft}
\begin{tabular}{llllllll}
\hline
\noalign{\smallskip}
RA ~ (2000) & DEC & Freq. &  Bandw. & R.Tel. (Config.)  &  Date  & Durat. & Figure\\
h ~ m ~ s & ~ $^{\circ}$ ~ $^{\prime}$ ~ $^{\prime\prime}$ &   MHz & MHz  & & & min \\
\noalign{\smallskip}
\hline
\noalign{\smallskip}
05 45 10.0 & $-$25 50 00 & 1365/1435 & 50 & VLA (C) & 28 Apr 00 & 110 & Figs.
\ref{vlaopt}, \ref{polar}, \ref{radx} \\
05 45 04.9 & $-$25 47 40 & 1385/1465 & 50 & VLA (BnA) & 18 May 02 & 20 &
Fig. \ref{relicS} \\ 
           &             & 4835/4885 & 50 &  VLA (BnA) & 18 May 02 & 20 & 
Fig. \ref{sourceA} \\ 
           &             & 8435/8485 & 50 &  VLA (BnA) & 18 May 02 & 20 \\ 
05 45 22.1 & $-$25 47 31 & 1385/1465 & 50 &  VLA (BnA) & 18 May 02 & 8 \\ 
           &             & 4835/4885 & 50 &  VLA  (BnA) & 18 May 02 & 8 \\ 
05 44 51.4 & $-$25 50 45 & 1384  & 104 &  ATCA (6A) &  17 Feb 00 & 720 
&  Fig. \ref{slee1} \\ 
           &            & 1384  & 104 &  ATCA (6B) &  22 Jun 00 & 720 &
Fig. \ref{slee1} \\  
           &            & 1704  & 104 &  ATCA (6B) &  22 Jun 00 & 720 \\
           &            & 2496  & 104 &  ATCA (6A) &  17 Feb 00 & 720 &
Fig. \ref{slee2} \\  
\noalign{\smallskip}
\hline
\noalign{\smallskip}
\end{tabular}
\end{flushleft}
\end{table*}

The presence of a relic source in A548 was suggested by Giovannini et
al. (1999), using radio data of the NRAO VLA Sky Survey (NVSS, Condon
et al. 1998).  The cluster A548, at an average redshift z=0.04
(Struble \& Rood 1999), shows a rather complex structure both in the
optical and in the X-ray band. Based on ROSAT data, Davis et
al. (1995) pointed out the existence of three main sub-clusters.
These are confirmed by
optical data which indicate also significant further substructures, some
of which overlap along the line of sight (Escalera et al. 1994, Den
Hartog \& Katgert 1996, Andreuzzi et al. 1998, Wegner et al. 1999,
Colless et al. 2001, Nikogossian 2001, Smith et al. 2004).  
The combination of X-ray and optical data  indicates  that A548
is a cluster in a collapsing phase and therefore not yet dynamically
relaxed.

The diffuse relic source detected by Giovannini et al. (1999)
by inspection of the NVSS is located in the subcluster A548b,
referred to in the literature also as A548S or A548W or A548SW,
at a redshift z= 0.0424 (Den Hartog \& Katgert 1996). 

Radio data for this cluster were  published previously by
Gregorini et al. (1994) and Marvel et al. (1999), but they refer to
radio galaxies only and Marvel et al's image does not cover the region
discussed in the present paper.

We present in this paper the results of new radio observations
obtained with the Very Large Array (VLA) and the Australia Telescope
Compact Array (ATCA) to study the radio emission in A548b.  The VLA
observations were aimed at imaging the diffuse emission with improved
sensitivity and resolution with respect to the NVSS.  The aim of the
ATCA observations was to search for fine structure in the image of the
radio relic similar to that seen in four relics by Slee et al. (2001),
and resolve possible discrete sources.  
Thanks to the high surface brightness sensitivity of the present 
radio data, we detect for the first time in a cluster
two radio relics located on the same side with respect to the cluster center.  
Moreover, a third possible diffuse source is detected close in projection to
the cluster center. The paper is organized as
follows: the radio data are presented in Sect. 2, the results from low
resolution images in Sect. 3, those from high resolution
images in Sect. 4. The results are discussed in Sect. 5.

For the computation of intrinsic parameters, we adopt 
the concordance cosmology with
H$_0$=70 km s$^{-1}$ Mpc$^{-1}$, $\Omega_m$ = 0.3, and $\Omega_{\Lambda}$ =
0.7. The luminosity distance D$_L$ of the subcluster under study is
188 Mpc; at this distance 1 arcsec corresponds to 0.84 kpc.

\begin{figure*}
\centerline{
\psfig{figure=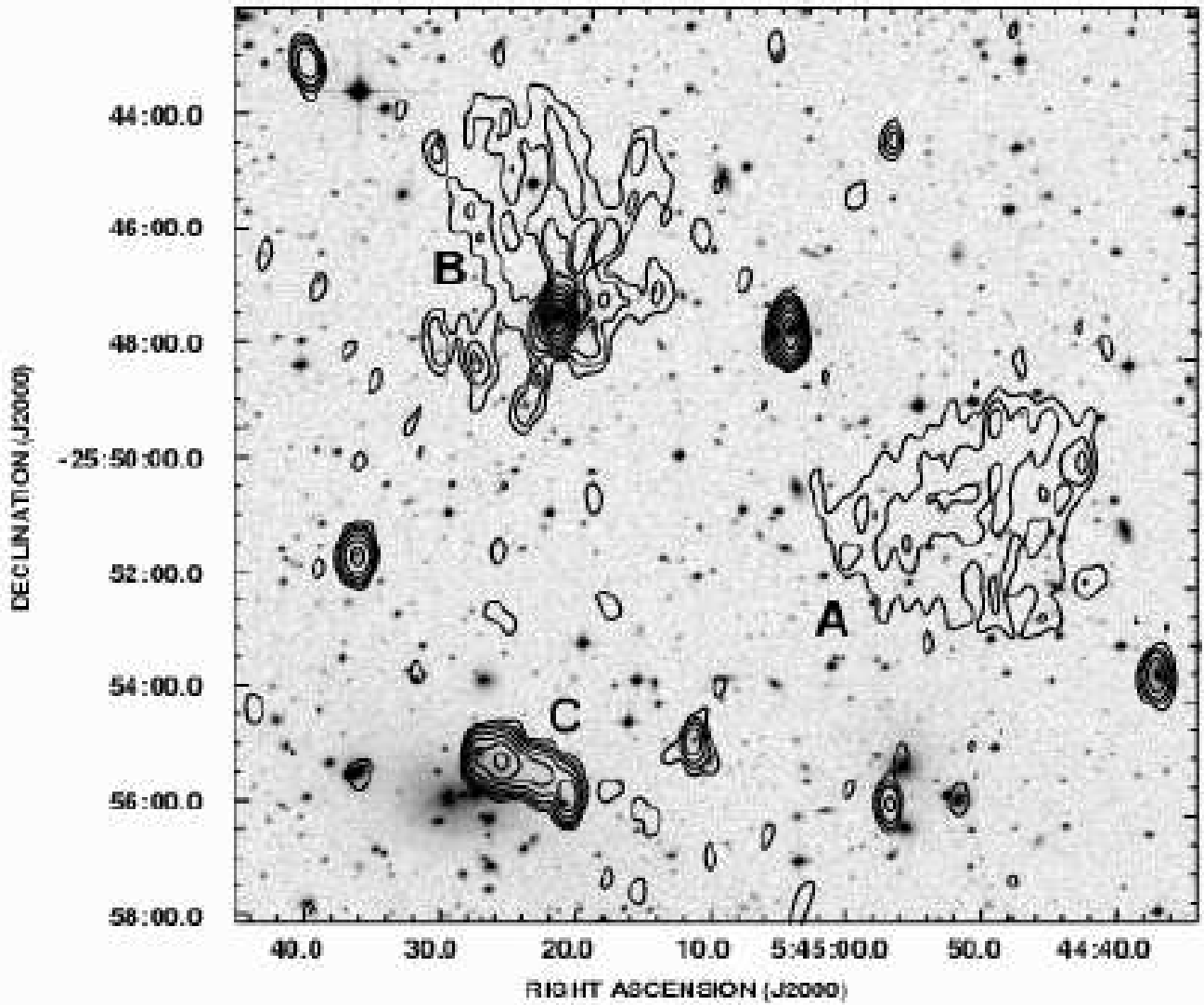,height=15cm}}
\caption{Contour radio 
emission in A548b at 1.4 GHz obtained with the 
VLA in C configuration at the resolution of  15$^{\prime\prime}$  $\times$ 30$^{\prime\prime}$
(FWHM, RA $\times$ DEC). Contour 
levels are  0.3, 0.6, 1, 2, 4, 8, 16, 32, 64 mJy/beam. 
The noise level is  0.09 mJy/beam. 
The grey-scale represents the optical R-band image taken from 
the DSS2. Labels A, B and C indicate diffuse radio sources.
} 
\label{vlaopt}
\end{figure*}

\section{Radio Observations}

\subsection {Very Large Array }

Radio observations at 1.4 GHz were obtained with the VLA in the C
configuration (see Table 1) in order to achieve good sensitivity to the
extended low-brightness structure.  The sources 3C48, 3C138 and
J0608$-$223 were observed as calibrators of the flux density scale,
the polarization position angle and the antenna gains and phases,
respectively.

Short exposures were obtained with the VLA in the BnA configuration at
1.4, 4.8 and 8.4 GHz centred on two unresolved sources in the region
of the diffuse emission, in order to obtain information on their
structure and spectra. The sources 3C48 and J0530$-$250 were used as
calibrators.

The data were calibrated and reduced with the Astronomical Image
Processing System (AIPS). Removal of bad data and interference was
done by editing and by means of the task VLALB, kindly provided by
W.D. Cotton.  Images were produced by Fourier-Transform, CLEAN
and RESTORE.  Several cycles of imaging and self-calibration were
performed, to minimize the effects of amplitude and phase variations.
The process started with a few iterations of phase calibration only,
then both phase and gain calibration solutions were obtained with a long
integration time, until no further 
improvement in the solutions was obtained. 
For the C configuration data, separate images for each of the
two observing frequencies were produced, in order to obtain an
estimate of the spectral index.  We also obtained an image of the
polarized intensity in the standard way.

\subsection {Australia Telescope Compact Array}

Radio observations of A548b were obtained at 1.4, 1.7 and 2.5 GHz with
the ATCA in two 6 km configurations; the observing sessions are
summarized in Table 1. The sources J1939$-$637 and J0608$-$223 were
used to calibrate the flux density scale and antenna gains and phases
respectively.

The data were edited, calibrated and reduced with the MIRIAD software
package, and the images were produced by following the standard
procedure of Fourier inversion (INVERT), CLEAN and RESTOR, combined
with two iterations of phase SELFCAL, followed by one
iteration of both phase and amplitude SELFCAL.

To search for fine structure in the image of the radio relic similar
to that seen in four relics by Slee et al. (2001), we needed to image
with the highest angular resolution available in the 1.4 and 2.5 GHz
bands. We were aware that the spatial frequencies available are not
low enough to reproduce a smooth image of several arcmin in angular
extent of the type detected by the NVSS at 1.4 GHz and reported by
Giovannini et al. (1999). However, if fine-scale structure with scale
size up to $\sim$ 1 arcmin were present, the 6A and 6B configurations
would be able to image it with reasonable sensitivity.

We did not attempt to combine the 1.4 GHz data from the different
VLA configurations and from ATCA because of the very different
pointing positions and bandwidths.  We note that the diffuse sources
are only detected on the short baselines of the VLA C configuration,
thus the combination of the uv-data would not improve the sensitivity
to these low surface brightness features. 

\begin{table*}
\caption{Diffuse radio sources in the field of A548b}
\begin{flushleft}
\begin{tabular}{lllllll}
\hline
\noalign{\smallskip}
 Name & RA ~ (2000) & DEC & S$_{1.4~GHz}$ & Largest Size & P$_{1.4~GHz}$ 
 & Fract. pol. \\
 & h ~ m ~ s & ~ $^{\circ}$ ~ $^{\prime}$ ~ $^{\prime\prime}$  &   mJy  &  arcsec ~ kpc  
& $\times$ 10$^{23}$ 
 W Hz$^{-1}$ & \% \\ 
\hline
\noalign{\smallskip}
A &    05 44 50 & $-$25 51 00 &  61$\pm$5 &   310 ~~~~  260  & 
   2.6$\pm$0.2  & $\sim$ 30 \\ 
B & 05 45 22 & $-$25 47 30 & 60$\pm$5 &  370  ~~~~   310   &   
2.5$\pm$0.2 & $\sim$ 30 \\
C & 05 45 26 & $-$25 55 16 & 77$\pm$6 &  160  ~~~~  135  &
  3.2$\pm$0.3 & $\sim$ 7 \\
\noalign{\smallskip}
\noalign{\smallskip}
\hline
\noalign{\smallskip}
\end{tabular}
\end{flushleft}
\end{table*}

\section{Results from low resolution images}

\begin{figure*}
\centerline{\psfig{figure=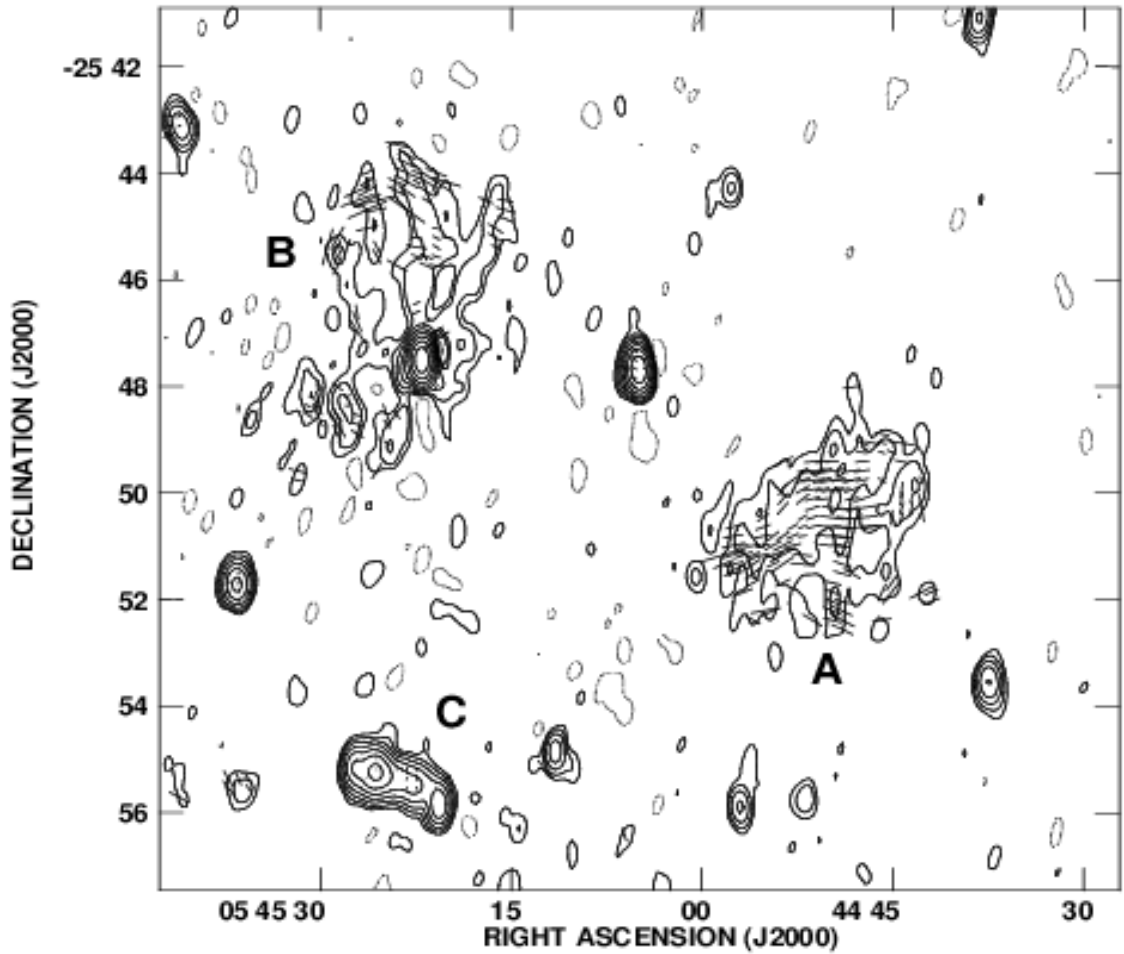,width=15cm}}
\caption{
Vectors of the polarized emission at 1.4 GHz superimposed 
on the total radio emission at the same frequency. Contour 
levels are  $-$0.25, 0.25, 0.5, 1, 2, 4, 8, 16, 32, 64 mJy/beam. 
The image 
of the polarized emission has been obtained
 with the VLA in C configuration with resolution of
15$^{\prime\prime}$ $\times$ 30$^{\prime\prime}$ (FWHM, RA $\times$ DEC) and  has a noise 
level of 0.04 mJy/beam. 
Vectors indicate the orientation of the electric field and
are proportional in length to the
fractional polarization with 1$^{\prime\prime}$~ corresponding
to 1.4\%. Labels A, B and C indicate diffuse radio sources.} 
\label{polar}
\end{figure*}

\begin{figure*}
\centerline{\psfig{figure=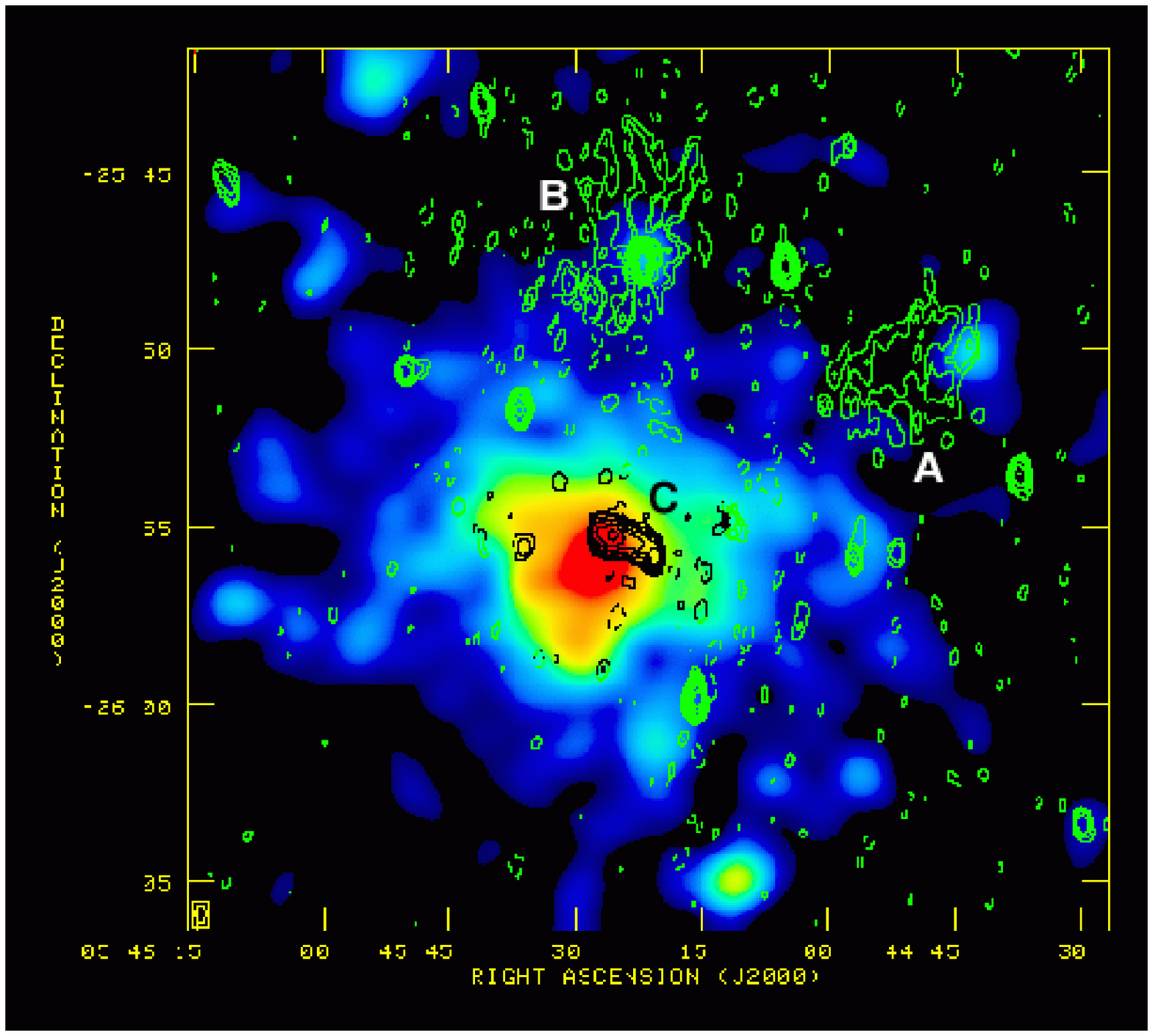,height=16cm}} 
\caption{Overlay of the radio image given in Fig. \ref{vlaopt}
(contours) onto the cluster  X-ray emission from ROSAT PSPC
(color-scale). Diffuse radio sources are labeled A, B and C, as
in Figs. \ref{vlaopt} and \ref{polar}.
}
\label{radx}
\end{figure*}

\subsection{Radio emission}

The radio image at 1.4 GHz, presented in Fig. \ref{vlaopt}, shows that
the diffuse radio emission in this cluster is quite complex.  A
diffuse source is detected in the western cluster region (A in
Fig. \ref{vlaopt}), confirming the emission seen in the NVSS.
This extended source has a quite regular morphology
and it is not associated with any galaxy.

Also, diffuse radio emission is detected in the northern cluster
region (B in Fig. \ref{vlaopt}) around a strong pointlike radio
source, in agreement with the traces of faint diffuse radio emission
in the NVSS image (Giovannini et al. 1999).  This extended emission is
filamentary and irregular. The embedded radio galaxy (ESO~488-G~006 
or PGC~17735) is not at the center of the diffuse radio emission, but
located toward its southern boundary.  The source parameters,
presented in Table 2, indicate that the two diffuse sources A and B
are strikingly similar in the total flux density at 1.4 GHz and
size. We also note that there is a  radio galaxy almost midway
between the two diffuse sources (see discussion in Sect. 4.3).

Another apparently diffuse source (labeled C in Fig. \ref{vlaopt}) is
detected $\sim$ 6$^{\prime}$~ south of the diffuse feature B.  It is
elongated, with no obvious optical identification, although it is
located \ltsim~1$^{\prime}$~ NW of the galaxy triplet VV 162, also
classified in the literature as a dumbbell, studied by Gregorini et
al. (1994).  Higher resolution images (see Sect. 4.4) are crucial to
disentangle the nature of this radio source.

We attempted an estimate of the spectral index\footnote{The spectral
index $\alpha$ is defined by S($\nu$) $\propto$ $\nu ^{\alpha}$} of
the two relics A and B, by comparing the images obtained from the two
separate VLA frequencies in L band, 1365 MHz and 1435 MHz. 
Although the frequencies are quite
close, there is indication of a steep spectrum with spectral index
$\sim$ $-$2 $\pm$ 1.

Polarized flux is detected in the diffuse radio sources at 1.4
GHz, as shown in Fig. \ref{polar}, where polarization vectors 
are overlayed onto the total intensity contours.  
The large relics (A and B) are polarized at a level of $\sim$ 30\%.
Polarized flux is also detected in
the central region of the source C, at a level of $\sim$ 7\%. No
information on the rotation measure can be obtained, thus no attempt
was made to correct the electric vectors for Faraday rotation.

\subsection{Radio versus X-ray comparison}

The X-ray emission of the cluster A548b has been analyzed by several
authors.  A temperature of kT= 3.1 $\pm$0.1 keV is obtained from ASCA
data (White 2000).  No cooling flow is reported at the cluster center
(White et al. 1997). The cluster 
vignetting-corrected X-ray surface brightness profile can
be represented by a $\beta$-model: 
$$S(r) =S_0 (1 + r^2/r^2_{\rm c})^{-3 \beta + 0.5}$$

\noindent
with $\beta$ = 0.52$^{+0.06}_{-0.04}$ and core radius r$_c$ =
115$^{+37}_{-29}$ arcsec (Neumann \& Arnaud 1999), corresponding to 
$\sim$ 96 kpc. 

\begin{figure*}
\centerline{\psfig{figure=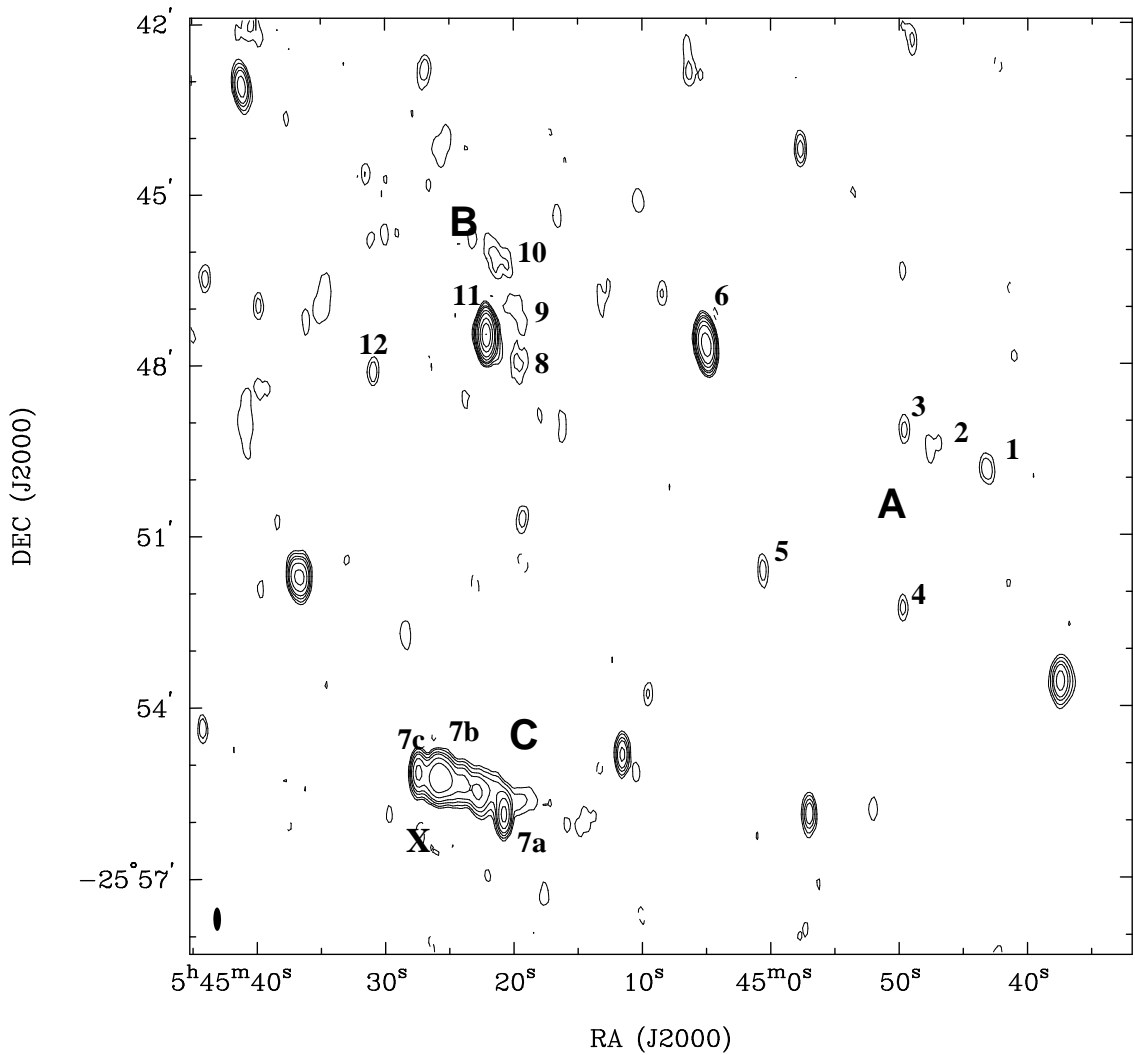,width=15cm}}
\caption{
The ATCA image of  A548b at 1384 MHz,
covering the same area as the VLA image of Fig. \ref{vlaopt}. 
Contour levels are
$-$0.2, 0.2, 0.4, 0.79, 1.32, 2.64, 5.28, 13.2, 26.4, 52.8 mJy/beam.
Sources of particular interest are numbered and mentioned in the 
text and in Tab. 3. 
The X symbol SE of the C-complex  indicates the X-ray centroid of the cluster
(see Sect. 3.2). The FWHM of the restoring beam is shown by the
filled ellipse in the bottom left corner and has dimensions 24.6$^{\prime\prime}$
$\times$ 8.0$^{\prime\prime}$~(PA = 0.5\degrees). The rms noise over the
image is 0.033 mJy/beam.
}
\label{slee1}
\end{figure*}

In order to compare the radio emission with the X-ray emission, we
have retrieved the X-ray image of  A548 from the ROSAT archive
(seq. number 800246). It was obtained with the Position Sensitive
Proportional Counter (PSPC) for a total exposure time of 10.9 ksec.
In Fig. \ref{radx} we show the overlay of the radio and X-ray images, for a
morphological comparison.  The radio image is the same as that
presented in Fig. \ref{vlaopt}.

The centroid that we derive for the X-ray emission of A548b is
consistent with the position given in the REFLEX catalogue RA$_{2000}$
= $05^h 45^m 27.2^s$, 
DEC$_{2000}$ = $-$25\degrees 56$^{\prime}$ 20$^{\prime\prime}$~
(B\"ohringer et al. 2004), and  with
the position of the dumbbell galaxy VV 162.
The two relics A and B are at projected distances of
$\sim$~500 and $\sim$ 430 kpc from the
cluster center and lie at the boundary of the X-ray brightness
distribution. They are well outside of  the cluster core radius,
but still well within the conventional Abell radius.

Source C  is clearly displaced from the cluster center and is 
located  about 1$^{\prime}$~ (corresponding to $\sim$ 50 kpc) due NW, 
i.e. within the cluster core radius (unless
strongly affected by projection effects).

\section{Results from high resolution images}

The two 1.4 GHz data sets from the ATCA 
(see Table 1) were concatenated to produce
the contour image of Fig. \ref{slee1}, which is a naturally weighted map
covering the same area as the VLA map of Fig. \ref{vlaopt}. 
The angular resolution of the ATCA map is 24.6$^{\prime\prime}$ $\times$
8.0$^{\prime\prime}$~(FWHM)  in PA = 0.5\degrees~
and the rms noise  after three iterations of SELFCAL
is 33 $\mu$Jy/beam.
Similar images were made at 1.7 and 2.5 GHz, but with somewhat
higher rms noise  of 52 and 41
$\mu$Jy/beam, respectively. All images
and the flux densities derived from them were corrected for primary
beam attenuation. The resolution achieved at 2.5 GHz  was 12.5$^{\prime\prime}$ 
$\times$ 4.2$^{\prime\prime}$~ 
in PA = $-$1.1\degrees. 

The 1.7 and 2.5 GHz images were useful in obtaining spectra for the
sources of interest in connection with the relics seen in
Fig. \ref{vlaopt}.  Information on individual radio sources is
also obtained from the VLA data at 1.4, 4.8 and 8.4 GHz in the BnA
configuration.  The list of discrete sources, together with their flux
densities, spectral indices and sizes, is given in Table 3, where flux
densities at 843 and 408 MHz are from Reynolds (1986). Errors of the
various parameters are given in parentheses.  The three regions A, B
and C (also No. 7), as well as the radio galaxy between the regions A
and B (No. 6), are discussed below.

\subsection {Region A}

The higher-resolution ATCA image (Fig. \ref{slee1}) shows no sign of the
smooth structure of relic A as seen in Fig. \ref{vlaopt}, but it shows
five fairly compact sources numbered 1, 2, 3, 4 and 5 within the
4.5$^{\prime}$ $\times$ 3.5$^{\prime}$~ area occupied by the relic. The same
sources are present in the 1.7 GHz image (not shown), enabling estimates
of their spectral indices (see Table 3).  These sources also appear as
peaks in the brightness contours of Fig. \ref{vlaopt}. None of these
sources can be identified with galaxy or stellar images on the
Digitized Sky Survey (DSS), e.g in SuperCosmos (see
http://www-wfau.roe.ac.uk/sss). The spectral indices of sources 1 and
2 are particularly highly negative, indicating that they probably
represent the peaks of fine structure present in the relic's
emission. Sources 3 and 4 have more normal spectral indices, although
still quite negative. Source 5 has a steep enough spectrum to again
associate it with fine structure in the relic's emission. It is not
surprising that these sources are not detected on the 2.5 GHz image as
their flux densities would fall below the 4$\sigma$ detection limit of
0.16 mJy/beam. We note that the spectral indices for sources 1, 2
and 5 are in a similar extreme range as those recorded by Slee et al
(2001) in the relics associated with A13, A85, A133 and A4038.

\subsection{Region B}

The 1.4 GHz ATCA image in Fig. \ref{slee1} shows that the relatively
strong source No. 11 (PMN J0545$-$2547) is surrounded on the western
side by more diffuse patches 8, 9 and 10, which are associated with
the more extended diffuse structure seen in the VLA image of
Fig. \ref{vlaopt}. The faint extension of source 11 to the SW, visible 
in Fig.  \ref{slee1}, is confirmed in the 1.7 GHz image.
Sources 8 and 9 appear in similar positions on the
1.7 GHz image, but there is no clear evidence for source 10.  These
diffuse patches do not appear to be associated with optical images on
any of the SuperCosmos plates.  Source No. 12 also appears to be
associated with the SE extension of the B-relic in Fig. \ref{vlaopt}.
However, it is coincident with an elliptical galaxy with m$_R$
$\sim$~18.6, so the source cannot be considered as part of the relic.

Source 11 is identified with the bright cluster galaxy ESO 488-G006
(PGC 17735, MCG-04-14-021, 2MASXi J0545220-254729) at z = 0.0385 and
J2000 position = 05$^h$ 45$^m$ 22.1$^s$, $-$25\degrees 47$^{\prime}$
30$^{\prime\prime}$.  It is classified as Sab (Dressler 1980) or S0 (ESO-Uppsala
Catalogue, Lauberts 1982) with m$_R$ $\sim$ 14. Its redshift is
consistent with it being a high-velocity member of A548b.  In the VLA
images at 1.4 GHz and 5 GHz it is pointlike at the resolution of about
1$^{\prime\prime}$. Based on the optical identification and its radio
structure, any connection to the diffuse emission B is unlikely.

The diffuse patches 8, 9 and 10 in Fig. \ref{slee1} are not well
fitted by elliptical Gaussians, and the details of their structure
differ from those in the 1.7 GHz image. Only their peak brightness
values are listed in Table 3, but their poor accuracy permits only
upper limits to be assigned to their brightness spectral indices. The
spectral index of 10 is particularly highly negative, indicating that
these three patches are very likely the brightest peaks in the fine
structure of the relic. Therefore we conclude that significant fine
structure with a  steep negative spectrum is present in the B-relic.
  
\subsection{Source 6}

The source No. 6 is located approximately midway between the two
relics A and B (Figs. \ref{vlaopt} and \ref{slee1}).  Thus, it might 
be the nucleus of a double-lobed radio galaxy,
whose extended lobes are the diffuse sources A and B.  We have
analyzed the structure of source  6, by means of high resolution
VLA images at different frequencies, to investigate if it could be
connected to the relics.

\begin{figure}
\centerline{\psfig{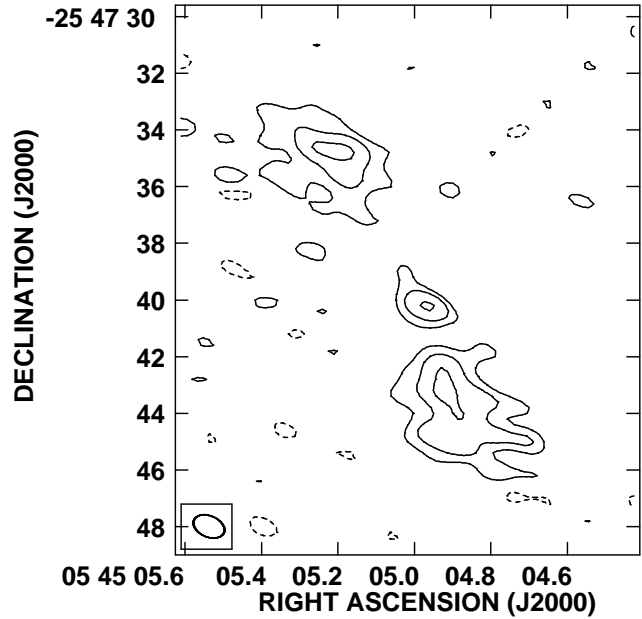}}
\caption{
Contour map of the radio galaxy No. 6 at 4.8 GHz, obtained
with the VLA in BnA configuration with resolution
of 1.2$^{\prime\prime}$$\times$0.7$^{\prime\prime}$. The noise level is 0.065 mJy/beam.
Contour levels are $-$0.15, 0.15, 0.3, 0.6 mJy/beam.
}
\label{sourceA}
\end{figure}

This source is identified with PGC 17721, a cluster galaxy at a
redshift = 0.0358, J2000 position = $05^h 45^m 04.9^s$, $-$25\degrees
47$^{\prime}$ 40.3$^{\prime\prime}$, classified as S03p (Dressler 1980).  Its
high-resolution radio structure obtained in C-band
(Fig. \ref{sourceA}) is that of a
classical double-lobed source, with a prominent nucleus, no clear jets and an
overall orientation in position angle $\sim$ 25\degrees (from N to
E).  This structure is confirmed by the image at 8.5 GHz (not shown).
Thus, this structure is oriented at a quite different position
angle with respect to the direction of  the sources A and B (about
60\degrees).
  
If the diffuse sources A and B were the extended lobes of a single
radio galaxy, the latter would have a total projected size of $\sim$ 650
kpc, and a total 1.4 GHz power of 3 $\times$ 10$^{23}$ W Hz$^{-1}$.
In the size-power diagram, this source would be among the largest
sources for that power. For this reason and for its high-resolution
structure, we consider the radio galaxy No. 6 as unrelated to the
diffuse sources A and B, and confirm the classification of the diffuse
features as relics.

\begin{figure}
\centerline{\psfig{figure=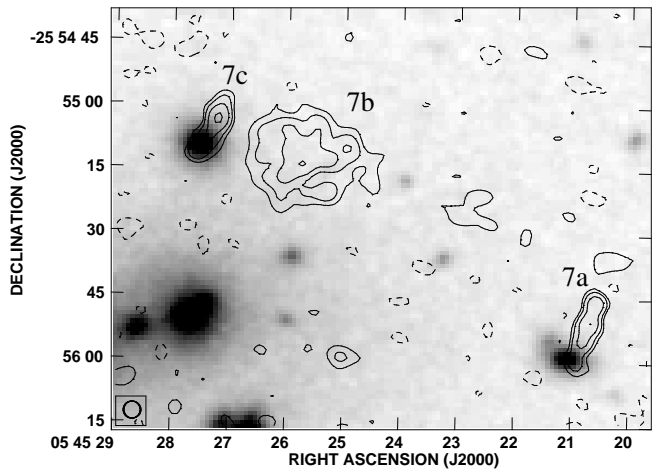,height=8cm}}
\caption{
Contour map of the extended source C (also 7) at 1.4 GHz, obtained with the VLA
in BnA configuration, at the resolution of 4$^{\prime\prime}$(FWHM). 
Contour levels: $-$0.2, 0.2, 0.4, 0.8, 1.6 mJy/beam. The noise 
level is 0.1 mJy/beam. The components 7a, 7b and 7c are discussed in the text.
The grey-scale represents the optical R-band
image taken from the  DSS2. }
\label{relicS}
\end{figure}

\begin{figure}
\centerline{\psfig{figure=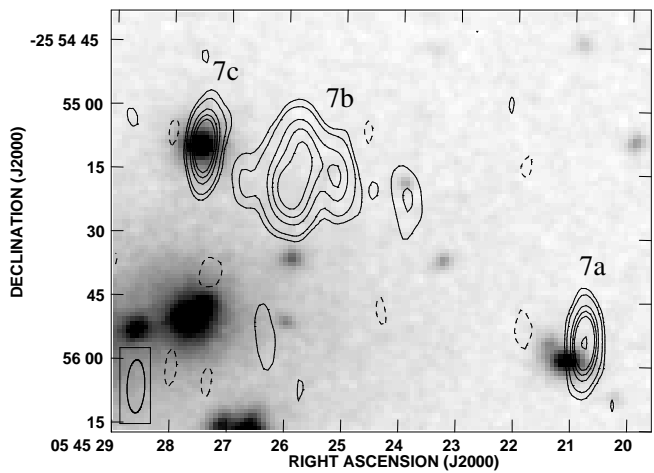,height=8cm}}
\caption{ The ATCA image of the extended source C (also 7) at 2.5 GHz,
covering the same area as the VLA image in Fig. \ref{relicS}.  The
grey-scale represents the optical R-band image taken from the
DSS2. Contour levels are $-$0.2, 0.2, 0.41, 1.02, 1.53, 2.04, 4.08
mJy/beam. The components 7a, 7b and 7c are discussed in the text.
The FWHM of the restoring beam has dimensions 12.5$^{\prime\prime}$ $\times$
4.2$^{\prime\prime}$ (PA = $-$1.1\degrees).  The rms noise in this peripheral
region of the imaged field is 0.075 mJy/beam.  }
\label{slee2}
\end{figure}

\subsection{Region C}

The diffuse source C (No. 7 in Fig. \ref{slee1}) consists of three
components 7a, 7b and 7c. This source complex lies within 1$^{\prime}$~ of
the X-ray emission centroid marked X in Fig. \ref{slee1} (Sect. 3.2),
and of the dumbbell galaxy.

In the higher-resolution VLA image (Fig. \ref{relicS}) at 1.4 GHz, a
central diffuse emission and two brighter and more compact spots are
visible.  The latter two are elongated in a radial direction with respect
to the map pointing centre (see Table 1), thus at least part of their
elongation is likely due to bandwidth smearing.  The total flux
density of this complex in the high-resolution image  (Tab.
3) is  only about 1/3 of that in the image at lower resolution (Tab. 2), thus
indicating the presence of diffuse structure of lower brightness.  We
note here that no sign of a radio source coincident with the brightest
cluster galaxy is detected.

The same structure is visible in the 2.5 GHz image shown in Fig.
\ref{slee2}. Due to the lack of sufficiently short 
baselines, the flux densities of the strongest component 7b at 1.4
and 2.5 GHz are seriously underestimated and no believable spectrum
can be fitted. Components 7a and 7c, with much smaller angular sizes,
yield more reasonable flux densities and spectral indices, which are
listed in Table 3.  The total flux in the three components is well
defined at all frequencies except 2.5 GHz, where the low flux density
is largely due to the low value for 7b.  The total flux measurements
yield a spectral index of $\alpha$ = $-$0.80 $\pm$ 0.06, which is more
typical of a radio galaxy than a relic and would not support the idea
that 7b, which contributes the major fraction of the total flux, is a
relic of the usual type. Perhaps the most unusual source of the trio
is 7c, which has a flat spectrum ($\alpha \sim$ 0).  The overlay to
the optical image (Figs. \ref{relicS} and \ref{slee2}) and reference
to NED reveal the presence of two bright galaxies in this complex.
The only galaxy that appears to be definitely associated with the
C-complex is 2MASXi J0545274-255509, which is a m$_R$ $\sim$ 15
elliptical within 1$^{\prime\prime}$~ from the position of
7c. Dressler \& Shectman (1988) report a redshift z = 0.0422 for this
galaxy, thus 7c is clearly identified with a galaxy in A548b. Also the
galaxy close to 7a is a member of the cluster, but is probably not
related to the radio source. Both the absence of an identification for
7b and its diffuse radio structure are at least consistent with its
status as a radio relic, despite the relatively normal spectral index
of the C complex.  Source 7b may also be a very distant radio galaxy
with an optical counterpart below the DSS plate limits, but we
consider this unlikely because of the source's amorphous morphology,
the lack of a compact nucleus and the missing flux at high resolution.
Moreover, a rather large size, of about 500 kpc, would be derived for
a radio galaxy at a redshift \gtsim~1.
Finally, we cannot exclude the possibility that source 7b 
is a lobe of either source 7a or 7c, however the lack of
a symmetric lobe located on the opposite side of the presumed nucleus
would lead to a quite unusual radio structure. 

\section{Discussion}

The radio data presented in this paper provide clear evidence
that both diffuse sources A and B are cluster relics. Indeed they 
are located at the cluster periphery, and they show
diffuse morphology, no connection to cluster galaxies, steep spectrum, 
and a high level of polarization.
The presence of two relics associated with the
intracluster medium (ICM) makes the cluster A548b very peculiar in the
radio domain.

This cluster is the first case where two relics are found
to be located on the same side with respect to the cluster center.
The two relics A and B show a roundish structure, and are about 300
kpc in size. They are highly polarized, as commonly found in cluster
relics (see e.g. Slee et al. 2001, Govoni \& Feretti 2004).  These
relics lie well within one Abell radius (2 Mpc with the adopted
cosmology), thus they cannot be classified as edge-located relics
unless they lie in the far front or back of the cluster.  In this
respect their projected distance from the cluster centre is similar to
that of the relic in A85,
discussed by Giovannini \& Feretti (2000) and Slee et al. (2001)
and modeled by En{\ss}lin \& Br\"uggen (2002). 
Spectral data on the A- and B-relics are presently very limited, 
with the only indication of $\alpha$ $\sim$ $-$2 from the
comparison of VLA images in L-band (Sect. 3.1).
The radio emission peaks numbered 1, 2 and 5 in
the A-relic have extremely high negative spectral indices between 1.4 GHz
and 2.5 GHz, indicating that fine structure is present over the
greater part of the relic. 
For the B-relic we can place an upper limit of $\alpha$ $<$ $-$2 on the
brightest peak (No. 10) in the fine structure.  

It is worth mentioning that, since the two relics are located on the same
side with respect to the cluster center, they could be connected to each
other, through a very low brightness emission.
The total size of this single relic would be about 650 kpc, and its
shape would be elongated roughly perpendicularly to the cluster
radius, as found in giant relics in cluster peripheral regions
(Giovannini \& Feretti 2004). The orientation of the magnetic field
that we observe (Fig. \ref{polar}) seems not aligned with the
elongated structure, as expected and detected in relics, but the lack
of information about Faraday rotation does not allow us to draw any
conclusion.

The status of the C-complex is very uncertain. It seems probable that
the more compact components 7a and 7c are not physically related to
the more extended 7b, although all lie in projection within $\sim$ 50
kpc of the X-ray centroid and the galaxy VV\,162.  The total 
flux measurements yield a
fairly normal spectral index indicative of a radio galaxy. If such is
the case, however, one would expect to detect a galaxy much brighter
than the DSS plate limits lying well within its contours. Thus the
C-complex might contain the brightest radio sources in a distant
cluster well beyond A548b. However, the amorphous morphology, the
absence of an active nucleus, and the presence of emission of very low
surface brightness as derived by the comparison between the total flux
density in the low- and high-resolution images (see Sect. 4.4) are in
favor of a diffuse nature for this complex, which we thus consider a
possible relic candidate. We remark that, because of the flux
missing in the high-resolution images, the diffuse emission
should extend further out of the structure labeled 7b, detected
in Figs. \ref{relicS} and \ref{slee2}.

 Small-size relics, although with quite steep spectra, have
been detected in A133 and A4038, located near the brightest cluster
galaxy (BCG), but not coincident with it.  For these sources, a
connection to the activity of the BCG is not clear (but see Fujita et
al. 2002 for A133).  It has been speculated (see e.g. Giovannini \&
Feretti 2004) that, if these diffuse sources would be old lobes of a
previous activity of the central galaxy, an almost symmetric double
structure centered on the galaxy should be detected in most cases.

Radio relics are detected in clusters both with and without a cooling
flow, suggesting that even minor or off-axis mergers may play an
important role in the external cluster regions.  Theoretical models
for the origin of relics propose that they are tracers of shock waves
in merger events (En{\ss}lin et al. 1998, En{\ss}lin \& Br\"uggen
2002).  A fraction of the energy dissipated in shock waves can be
transferred to relativistic particle populations. Accelerated
relativistic electrons have short radiative lifetimes and should
therefore produce the radio relic emission close to the current
location of the shock waves.

The clusters of the A548 complex are likely in a state of interaction,
thus the presence of relics in A548b confirms the association between
relics and mergers. In this framework, the relics A and B would trace
a shock wave in an outer cluster region, whereas the source C, if 
confirmed as a relic, 
would indicate a much more internal shock wave. 
In this respect, this cluster offers a unique scenario
to investigate the evolution of a cluster merger.

A deeper inspection of the radio images in both Figs. \ref{vlaopt} and
 \ref{slee1} showed that the majority of compact radio sources even
 down to the faintest radio flux levels (\gtsim~0.25\,mJy, or $P_{\rm
 1.4\,GHz}$ \gtsim~$10^{21}\,W\,Hz^{-1}$ at a redshift of $z$=0.0424)
 have optical counterparts. We plan to discuss this fact and its
 implications in a later paper.

\section{Conclusions}

From the results and discussion presented in this paper
we derive the following conclusions.

Extensive radio observations of the two diffuse
radio sources A and B in the the cluster of galaxies Abell 548b confirm
their classification as radio relics. The two relics, of similar flux
density, extent, shape, polarization and spectral index, are located
at projected distances of about 430 and 500 kpc from the cluster
center, on the same side of the cluster's X-ray peak.

We present images at high resolution of the cluster and measure
radio spectral indices of discrete radio sources embedded within the
diffuse features.  We attempt to distinguish emission peaks of the
diffuse sources from unrelated sources, and we find that both relics
show possible fine structure, when observed at high resolution.
We analyze the  high resolution images of the strong radio source 
located approximately midway between the two relics, and 
conclude that it has very likely no physical connection to the
diffuse sources. 

Another diffuse source C is detected close in projection to the
cluster center.  High-resolution images show that it contains two
discrete radio sources 7a and 7c and a diffuse component 7b.  
The nature of this
complex is uncertain. It seems probable that the more compact
components are not physically related to the more extended one,
although all lie in projection within $\sim$ 50 kpc of the X-ray
centroid and the galaxy VV\,162.  Although the spectral index of the
complex is fairly normal, the amorphous morphology of the extended
component, the absence of an active
nucleus, and the missing flux in the high-resolution images, are in
favor of a diffuse nature for this complex, which we thus consider a
possible relic candidate.

We suggest that the two external relics may trace the same shock
wave in the outer cluster region, whereas the possible relic close to
the cluster center would indicate a much more internal shock wave.
Future studies in the radio and X-ray domains will allow to detect the
details of the merger process and shed more light on the connection
between cluster mergers and the formation of radio relics.

\section*{Acknowledgments}

We thank Loretta Gregorini for kindly supplying the 4.8 GHz image
of the central cluster region. We are indebted to B. Komberg for valuable
comments.
We acknowledge the anonymous referee for useful suggestions.

This work has been partially supported by the Italian Ministry 
for University and Research (MIUR) under grant PRIN 2004029524-002. 
H.A. acknowledges support from CONACyT grant 40094-F.

This research has made use of the NASA/IPAC Extragalactic Database
(NED) which is operated by the Jet Propulsion Laboratory, California
Institute of Technology, under contract with the National Aeronautics
and Space Administration. We also made use of the database
 CATS ({\tt cats.sao.ru}, see
Verkhodanov et al., 1997) of the Special Astrophysical Observatory.

The National Radio Astronomy Observatory is operated by Associated
Universities, Inc., under contract with the National Science
Foundation.

The Compact Array is part of the Australia
Telescope, which is funded by the Commonwealth of Australia for
operation as a National Facility managed by CSIRO.

\newpage

\begin{figure*}
\centerline{
\psfig{figure=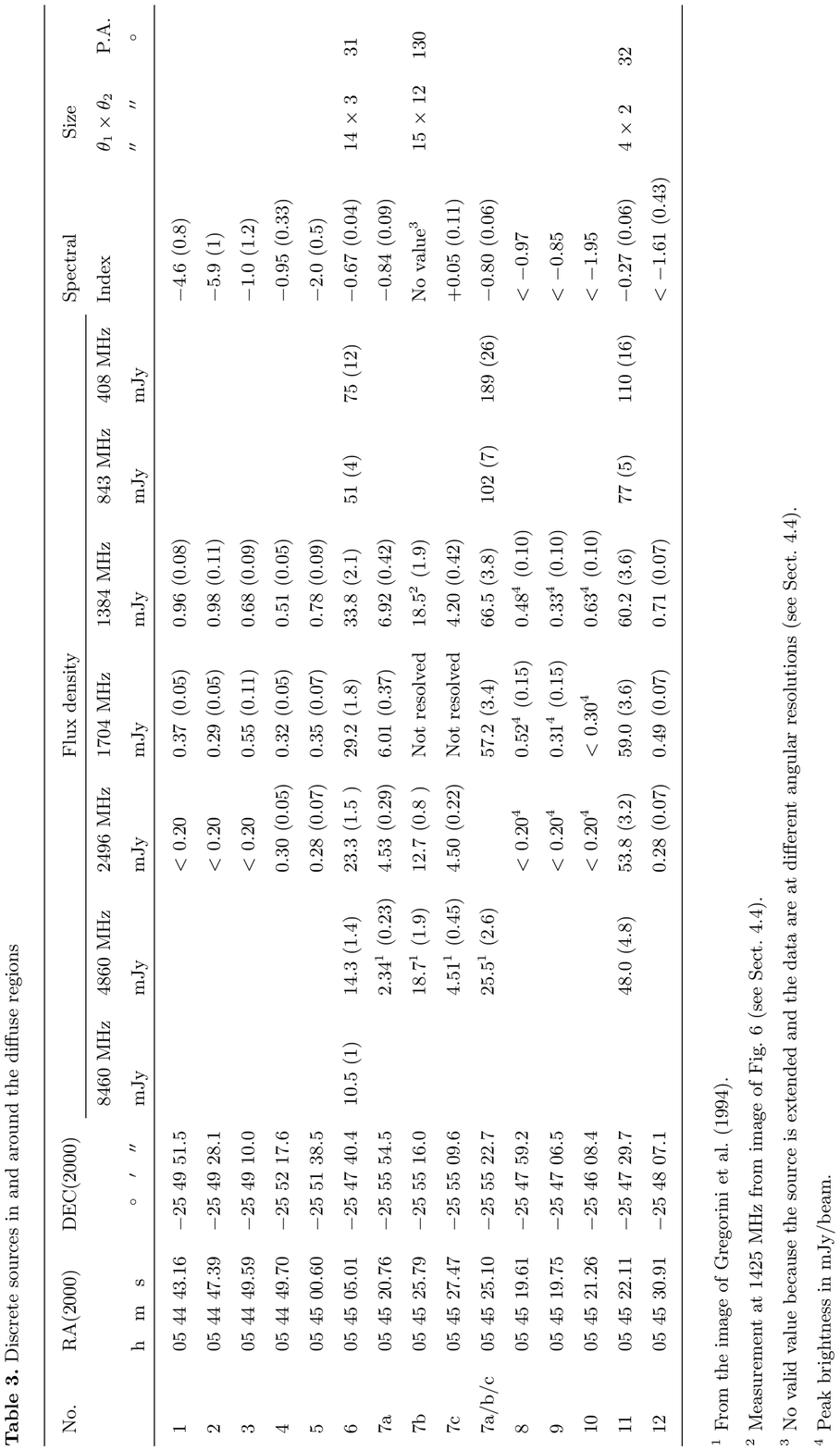,width=13cm}}
\end{figure*}

\end{document}